\documentclass[galaxies,article,accept,pdftex,moreauthors]{Definitions/mdpi} 

\firstpage{1} 
\pubvolume{1}
\issuenum{1}
\articlenumber{0}
\pubyear{2024}
\copyrightyear{2024}
\externaleditor{Academic Editor: Oleg Malkov}
\datereceived{8 May 2024} 
\daterevised{22 July 2024} 
\dateaccepted{2 August 2024} 
\datepublished{5 August 2024} 
\hreflink{https://doi.org/10.3390/galaxies12040044} 

\Title{A Dust-Scattering Model for M1-92: A Revised Estimate of the Mass Distribution and Inclination}
\TitleCitation{A Dust-Scattering Model for M1-92: A Revised Estimate of the Mass Distribution and Inclination}

\Author{Yun Qi Li 
 $^{1,}$*
\orcidA{}, Mark R. Morris $^{1,}$*\orcidB{} and Raghvendra Sahai $^{2}$\orcidC{}}

\AuthorNames{Yun Qi Li, Mark R. Morris and Raghvendra Sahai}

\AuthorCitation{Li, Y.Q.; Morris, M.R.; Sahai, R.}

\address{%
$^{1}$ \quad  Department of Physics \& Astronomy, University of California, Los Angeles, Los Angeles, CA 90095-1547, USA\\
$^{2}$ \quad Jet Propulsion Laboratory, California Institute of Technology, Pasadena, CA 91109, USA; sahai@jpl.nasa.gov 
}
\corres{Correspondence: yunqil@ucla.edu (Y.Q.L.); morris@astro.ucla.edu (M.R.M.)}

\abstract{Preplanetary nebulae (PPNe) are formed from mass-ejecting late-stage AGB stars. Much of the light from the star gets scattered or absorbed by dust particles, giving rise to the observed reflection nebula seen at visible and near-IR wavelengths. Precursors to planetary nebulae (PNe), PPNe generally have not yet undergone any ionization by UV radiation from the still-buried stellar core. Bipolar PPNe are a common form of observed PPNe. This study lays the groundwork for future dynamical studies by reconstructing the dust density distribution of a particularly symmetric bipolar PPN, M1-92 (Minkowski's Footprint, IRAS 19343$+$2926). For this purpose, we develop an efficient single-scattering radiative transfer model with corrections for double-scattering. Using a V-band image from the Hubble Space Telescope (HST), we infer the dust density profile and orientation of M1-92. These results indicate that M1-92's slowly expanding equatorial torus exhibits an outer radial cutoff in its density, which implicates the influence of a binary companion during the formation of the nebula.} 

\keyword{preplanetary nebulae; M1-92 (Minkowski's footprint); post-AGB stars; mass loss; \linebreak circumstellar matter morphology}

\begin{document}

\section{Introduction} \label{introduction}

Preplanetary 
nebulae (PPNe) arise in the last stages of the transition from the asymptotic giant branch (AGB) to the planetary nebula (PN) stage of stellar evolution \citep{sahai_preplanetary_2007}. They are often prominent reflection nebulae, as~viewed in the optical and near-infrared, as~a result of the scattering of light from the central star off dust particles that recently formed in the outflowing material ejected by the rapidly evolving central star. PPNe are generally distinguished from planetary nebulae by the absence of a central ionized region that has been created by the exposure of a UV-emitting stellar~core. 

Bipolar morphologies are very commonplace among preplanetary nebulae. They typically consist of two axisymmetric lobes (sometimes point-reflection-symmetric), diametrically placed on opposite sides of the central star \citep{morris_bipolar_1990, soker_jet_1992, mastrodemos_bipolar_1998}. The~bipolar geometry is attributable to the systemic latitude dependence of both the outflow velocity and the mass-ejection rate, the~combination of which determines the local density of the scattering dust particles at any given radius and latitude. Because~the dust density is typically large near the system's equatorial plane, the~opacity there is large, and little scattered light can escape at smaller latitudes, but~at higher latitudes, where the dust and gas are more rarified, the~starlight can more easily penetrate and therefore be scattered. Consequently, in~many of the most prominent cases, the~light emerges from the bipolar lobes, where there is a relative paucity of gas and dust. In~cases where the mass loss rate is low, however, the~equatorial concentration of dust is not as optically thick and the starlight can be scattered primarily off material in that equatorial concentration, forming what appears in scattered light to be an expanding torus or toroid \citep{sahai_preplanetary_2007}. 

The latitude dependence of the mass loss rate in bipolar PPNe has been subject to debate but~is often attributed to binary interactions. The~gravitational effects of a close companion favor mass loss at low systemic latitudes \citep{morris_mechanisms_1987, livio_common_1988, taam_double-core_1989, morris_optical_1990, livio_common_1990, livio_common_1992, sahai_multipolar_1998, mastrodemos_bipolar_1999, de_marco_origin_2009}. Additionally, angular momentum transfer from a close binary interaction is most effective at low systemic latitudes and can be invoked to expel mass equatorially at a fairly high rate \citep{morris_bipolar_1990}. In~the most extreme cases, high mass-loss rates are produced during the transition to a common-envelope configuration, a~process that happens with a sudden onset on a very short timescale \citep{livio_common_1988, morris_bipolar_1990, morris_optical_1990, soker_formation_1990, soker_jet_1992, mastrodemos_bipolar_1998, soker_binary_1998, de_marco_origin_2009, ivanova_common_2013, shiber_companion-launched_2019, zou_bipolar_2020}. This process occurs before the bipolar collimated fast wind that creates the lobes is initiated \citep{morris_bipolar_1990}. 

\begin{figure}[H]
  
  \includegraphics[width = 5in]{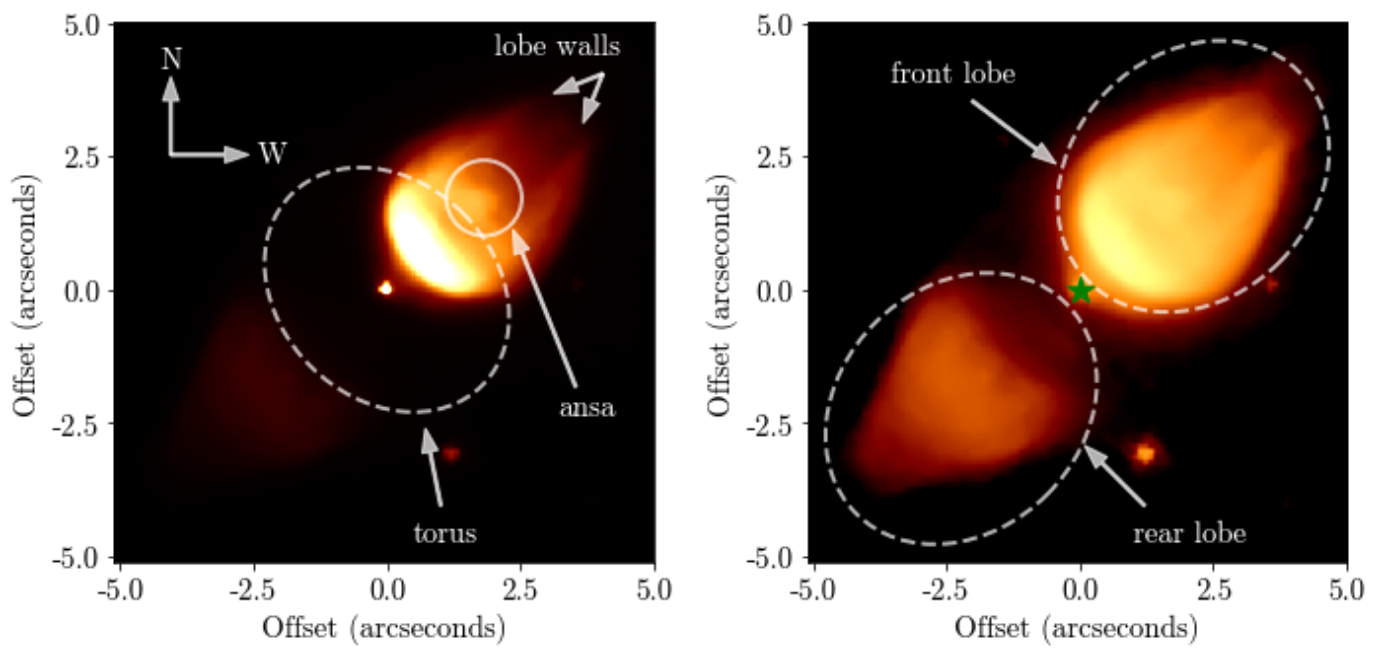}
  
    \caption{Hubble 
 Space Telescope (HST) image of M1-92, observed with the F547M filter (central wavelength $5483.0\textup{~\AA}$ and bandwidth $483.0\textup{~\AA}$) and taken by the Wide Field and Planetary Camera 2 (WFPC2), using aperture pc1 with 80 s of exposure. \textbf{Left}: linear scale. The~silhouette of the optically thick torus is represented by an elliptical outline. The~front lobe ansa is shown within the circle. The~lobe walls, which are brighter than the interior of the lobes, can best be seen at the tip of the front lobe. In~this panel, the~color map ranges from 0 to 15 counts per second. 
    \textbf{Right}: logarithmic scale. The~rear lobe is visible with this scaling. The~star's position is highlighted with the green symbol. In~this panel, the~color map spans a range of 0.1 to 100 counts per second. 
    This image was taken on \linebreak 30 May 1996 \citep{bujarrabal_shock_1997}. An~RGB figure of M1-92 with line emissions that clearly display the ansae can be found in Figure~1 of 
 \citet{balick_models_2020}.}
    \label{fig:hst-minkowski}
\end{figure}

We here report our investigation of a typical bipolar preplanetary nebula, M1-92, which is estimated to lie at a distance of 2.5 kpc \citep{feibelman_ultraviolet_1990, bujarrabal_shock_1997}. {M1-92 has one of the best observed axial symmetries, which is evident in Figure~\ref{fig:hst-minkowski}. The~bipolar outflow of M1-92 has been underway for approximately 1000 years \citep{alcolea_minkowskis_2007, alcolea_new_2008, alcolea_m_2022}, giving rise to a nebula that extends about 12$''$\linebreak  (30,000 AU) tip-to-tip. M1-92 is a typical ``high mass loss rate'' bipolar PPN, where there is an optically thick equatorial dust concentration. Close to the star, the~dust concentration forms a thin, flat, expanding disk that flares outward at greater distances \citep{alcolea_minkowskis_2007}. We refer to this concentration as the ``torus''. Most of the visible light comes from starlight scattered off dust at higher latitudes, within~two collimated, high-velocity outflows in diametrically opposite directions (the lobes). In~this paper, we present and describe a dust scattering model that reproduces the optical (V-band/F547M) appearance of M1-92. The~model-inferred dust morphology offers insights into M1-92's formation and constrains the evolution of the nebula.}

In the observed optical images of M1-92, the~rear lobe vanishes abruptly at radii below a fixed radius to the central star; \citet{morris_prevalence_2015} suggested that this was a result of M1-92's optically thick torus being sharply terminated at a radius that projects to the observed cutoff value. They attributed this characteristic to the onset of strong mass loss provoked by a close binary interaction, possibly the immersion event leading to a common envelope binary. {This characteristic is observed in several other PPNe as well, but~it is most clearly seen in the optical images of M1-92.} Therefore, a~crucial objective of our model is to explore the extent to which the density of the torus abruptly declines by~modeling the torus' absorption effect on the rear lobe of M1-92. {The precise geometric} modeling of the sharp radial cutoff of the expanding torus is essential for determining whether a sudden mass-loss event, such as that which would occur when a binary system enters the common envelope phase, is necessary to explain the torus'~formation.



\section{Data}

Five Hubble Space Telescope (HST) WFPC2 F547M (central wavelength $5483.0\textup{~\AA}$, bandwidth $483.0\textup{~\AA}$) observations of M1-92 are available in the Barbara A. Mikulski Archive for Space Telescopes (MAST). One image was acquired by PI Susan R. Trammell on 4 April 1996 \citep{trammell_hubble_1996}, using aperture pc1-fix with 350 s of exposure. Four additional images were acquired by PI Valentin Bujarrabal on 30 May 1996 \citep{bujarrabal_shock_1997}. These images are observed using the aperture pc1. Two of these four May 30 observations were exposed for 400 s, which saturates the central star. The~other two have exposure times of 80 s, for~which the peak brightness of the central star are 2870 and 2963 DN, respectively, less than WFPC2's practical saturation limit of 3500 DN \citep{gilliland_stellar_1994, mcmaster_hubble_2008}. Therefore, these two images are suitable for quantitative analysis involving the brightness of the central star. Figure~\ref{fig:hst-minkowski} shows the first 80 s exposure~image. 

Per \citet{bujarrabal_shock_1997}, standard reduction procedures were applied to their observations. Different exposures were rotated to align with each other, averaged, and~finally oriented with the  north side facing up. We used both pre- and post-reduction images. Sampling during the rotation process of the reduction causes the resolution of the images to be reduced by a factor of $2.2$ \citep{bujarrabal_shock_1997}. We therefore used pre-reduction images for determining the point spread function and post-reduction images for displaying resolved structures within the~nebula. 

\section{Methods} \label{methods}

\begin{figure}[H]

\begin{adjustwidth}{-\extralength}{0cm}
\centering 
 \includegraphics[width = 7in]{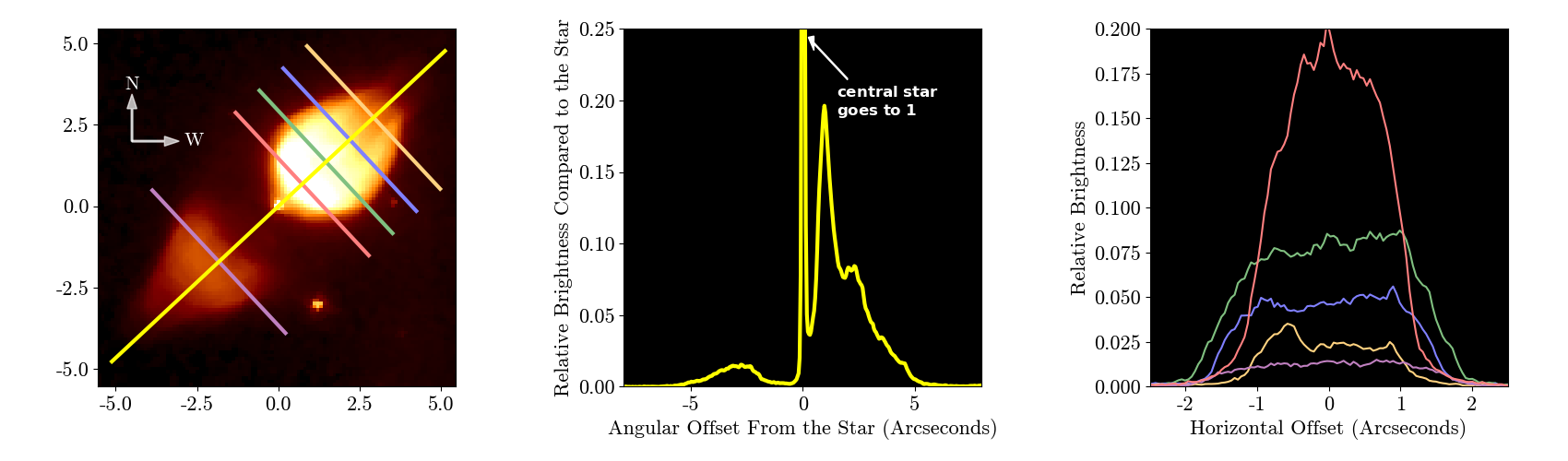}
\end{adjustwidth}
  
    \caption{Hubble 
 Space 
 Telescope (HST) image and extracted light profiles of M1-92, using an \mbox{80 s} exposure. \textbf{Left}: Observed image in square-root scale and placement of the profiles. The~yellow profile is along the system's symmetry axis, while the orange, blue, green, red, and~purple profiles are perpendicular to the symmetry axis and~are offset by 4$''$, 3$''$, 2$''$, 1$''$, and~$-$2.5$''$ from the star. The~color map spans a range from 0 to 10 counts per second.
    \textbf{Middle}: Normalized axial brightness profile of M1-92 in yellow, with~the peak brightness of the star set to a value of 1. \textbf{Right}: a plot of the five horizontal profiles. }
    \label{fig:hst-profiles}
\end{figure}

Our modeling is {constrained by} the spatial distribution of the intensity in the observed V-band/F547M image from input morphological parameters using a single-scattering radiative-transfer calculation. Our goal is to constrain the values of the input parameters of the 3-dimensional model that best reproduces the measured 2-dimensional image. The~goodness of the fit to the image is quantified by using a set of 1-dimensional profiles through the image, as~shown in Figure~\ref{fig:hst-profiles}. 
\vspace{-6pt}
\unskip

\subsection{Dust~Distribution} \label{dust-distribution}

Despite M1-92's apparent axial and reflection symmetry, its morphology is complex. Observed structures that the model must account for include the torus, the~bipolar lobes, the~high-density shells around the lobes (the “lobe walls”) \citep{bujarrabal_direct_1994, bujarrabal_shock_1997, ramos-medina_spatio-kinematics_2014, alcolea_m_2022}, and~the ansae, which are optically bright knots positioned along the system's symmetry axis at/near the middle of the lobes \citep{soker_formation_1990, morris_bipolar_1990, trammell_hubble_1996}. \citet{alcolea_minkowskis_2007, alcolea_m_2022} observed molecular emission from the ``polar tips'' at the two ends of the lobes. Since they are too far out to contribute noticeably to optical emission, we did not include them in the~model. 

A large number of independent input morphological parameters are required to construct the included structures, while additional structures/refinement parameters can also be raised. We continually verify the necessity of each parameter, removing those that are neither optically constrained nor substantiated by other research (e.g., the ellipticity of the torus and the distance dependence of the density in the lobe walls). Most of the free parameters are those that specify the dust density as a function of latitude, radius, and the inclination of the system's symmetry axis. Table~\ref{tab:strongly-constrained-parameters} contains the four strongly constrained parameters, while Tables~\ref{tab:parameter-fixed} and~\ref{tab:parameter-best} in Appendix \ref{model-parameters} list all input parameters and their~values.

\renewcommand{\arraystretch}{2}
\begin{table}[H]
\caption{Strongly 
 constrained~parameters.}
\label{tab:strongly-constrained-parameters}
\setlength{\tabcolsep}{3.85mm}

\begin{tabular}{l l}
\toprule
\textbf{Parameter Name} & \textbf{Parameter Value} \\ 
\midrule
Total Gas/Dust Mass & \(M_{tot}^*(\theta=66^\circ)=1.10 \pm 0.1\;M_\odot \cdot \left(\frac{d}{2.5\text{kpc}}\right)^2 \cdot \left(\frac{\text{GDR}}{200}\right)\) \\
\multicolumn{2}{l}{\quad torus: $1.02\;M_\odot$} \\
\multicolumn{2}{l}{\quad lobe walls: $7.2\times10^{-2}\;M_\odot$} \\
\multicolumn{2}{l}{\quad evacuated lobe interiors: $8.6\times10^{-3}\;M_\odot$} \\
\multicolumn{2}{l}{\quad ansae (each): $2.7\times10^{-4}\;M_\odot$} \\
Inclination & \(\theta^*(g=0.6)=66^\circ \pm 3^\circ\) \\
Torus cutoff radius & \(R_{out}^*=6685 \pm 600\;\text{AU} \cdot \left(\frac{d}{2.5\text{kpc}}\right)\) \\
Sharpness of the torus cutoff & \(dr_{out}^*(\phi=0^\circ)/R_{out}^* \leq 0.2\) \\[1ex]
\bottomrule
\end{tabular}
\end{table}

Given this high-dimensional modeling task, it is not practical to investigate the parameter space numerically in order to simultaneously find the absolute best fit for all parameters. Therefore, we have manually optimized the fit by iteratively adjusting one parameter at a time. However, optimized values of important parameters listed in \linebreak Table~\ref{tab:strongly-constrained-parameters} are found with an algorithm-based optimization procedure detailed in \mbox{Appendix \ref{app:parameter-optimization}}. The~semi-manual adjustment process allows us to arrive at plausible parameter values more expeditiously than would be feasible with a high-dimensional global minimization algorithm and~permits us to also steer away from those regions of parameter space that give rise to models that are not hydrodynamically realistic. Our model fully utilizes the symmetries of M1-92 to facilitate an efficient parameter~search. 

For the dust grains in the model, we adopt the Mathis--Rumpl--Nordsieck size distribution: $n(a)\propto a^{-3.5}$ \citep{mathis_size_1977, weingartner_dust_2001, murakawa_evidence_2010}. We follow \citet{murakawa_evidence_2010} in using a minimum grain size of $5\times10^{-3}\;\upmu$m, a~maximum grain size of $1000\;\upmu$m in the torus, and~a maximum grain size of $0.5\;\upmu$m everywhere else. The~size-dependent albedo is determined based on the scattering-efficiency and extinction-efficiency functions from~\cite{draine_physics_2011}. Similar to \citet{yusef-zadeh_bipolar_1984}, we calculate the angular-dependence of scattering using the Henyey--Greenstein phase function \citep{henyey_diffuse_1941}, with~a {scattering asymmetry factor of $g=\langle \cos{\theta} \rangle=0.6$, derived both from simulated results \citep{weingartner_dust_2001, draine_interstellar_2003} and from observations \citep{witt_scattering_1982, witt_scattering_1990}. Since the scattering asymmetry factor $g$ is equal to the mean cosine of the scattering angle, a~higher $g$ indicates stronger forward scattering. $g=0$ indicates isotropic scattering.} We assume a gas/dust ratio of 200 \citep{sahai_quadrupolar_2007}.

{Our models are fitting for the density across the nebular medium, which is not based on a hydrodynamic calculation. Density within the torus and the lobe interiors are specified using the continuity equation:}
\begin{equation}
    \rho (r,\phi)=\frac{\dot{M}(r,\phi)}{\hat{A}(r,\phi) V(r,\phi)},
    \label{eqn:mass-density}
\end{equation}
{where $\phi$ is the systemic latitude, $\dot{M}$ is the mass loss rate, $V$ is the outflow velocity, and~$\hat{A}$ is an estimate of the cross-sectional area of the outflow (the torus and the lobes are treated as different outflows). $\dot{M}$ and $V$ can be both latitude- and radius-dependent depending on the nebular structure as well as on temporal variations in the outflow. The~value for $\hat{A}(r,\phi)$ is different for the torus and the lobe interiors. For~the torus, we use $\hat{A}(r,\phi)\propto r^2$ for $r\geq 1.8\times10^{16}$ cm (0.5$''$), and~$\hat{A}(r,\phi)\propto r$ for $r< 1.8\times10^{16}$ cm. For~the tenuous lobe interiors, we set $\hat{A}(r,\phi)$ to be the cross-sectional area of the lobe interior at $r$, assuming that the lobe outflow is confined by the lobe walls and that no mass is exchanged between the interior of the lobes and the lobe walls\endnote{These assumptions are not crucial as the lobe interiors contribute little to the scattered light, as~described in Section~\ref{discussion}.}. The~density within the lobe interiors is thus dependent on the geometrical shape of the lobes.}

\textls[-15]{{Since density is specified using both $\dot{M}(r,\phi)$ and $V(r,\phi)$, identical model-constrained density distributions can be reproduced by different combinations of mass loss rates and outflow velocities. Expansion velocities within M1-92 are well studied; therefore, we set them as fixed parameters, while the mass outflow rates at each latitude are free parameters that can be inferred from the constrained density.} In the model-optimization procedure, we referred to velocity measurements from \citet{seaquist_oh_1991, trammell_spectropolarimetry_1993, bujarrabal_direct_1994, trammell_hubble_1996, bujarrabal_shock_1997, arrieta_protoplanetary_2005, alcolea_minkowskis_2007, ueta_hubble_2007, ramos-medina_spatio-kinematics_2014, alcolea_m_2018, alcolea_m_2022}. {We employ radius-independent outflow velocity values of $50$ km s$^{-1}$ in the lobe interiors and $8$ km s$^{-1}$ in the torus \citep{bujarrabal_structure_1998, alcolea_minkowskis_2007}.}}

The densities in the lobe walls and ansae are specified by constant density laws over both radial distance and latitude since the formation of these structures is not likely the result of a purely radial outflow but~rather by a compression of the torus by the ram pressure of the high-velocity outflow in the lobes. We tested several models for the density of the lobe walls and found a constant density law provides the best-fitting~model. 

\begin{figure}[H]

\begin{adjustwidth}{-\extralength}{0cm}
\centering 
    \includegraphics[width = 6in]{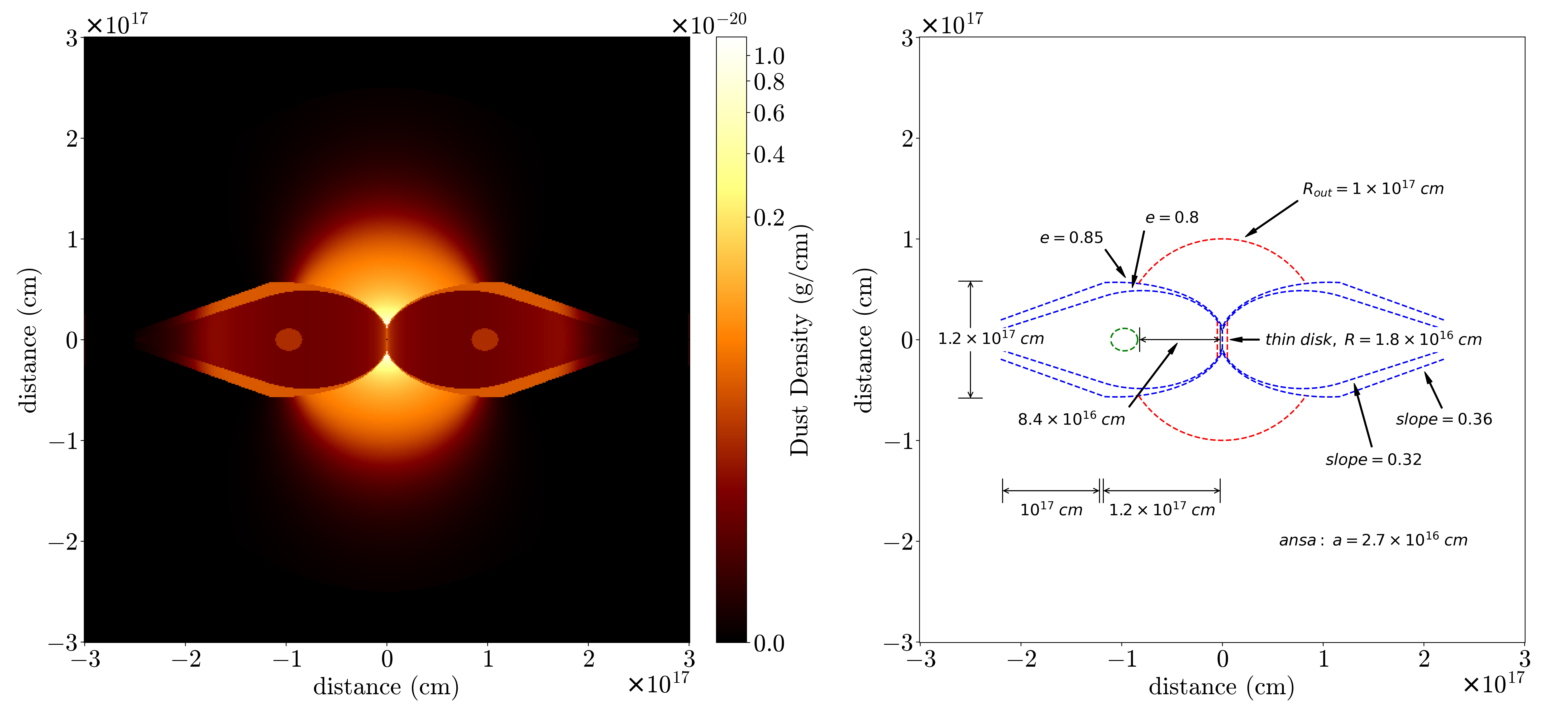}
\end{adjustwidth}
    \caption{{The}  best-fitting dust density map inferred from the optical image. \textbf{Left}: the density distribution in a meridional cross-section, with~values scaled to the $0.2$ power to best illustrate the density distribution at larger radii. \textbf{Right
}: the geometry of each component included in the presented density map. Note the axial and reflection symmetries, the~ansae within the lobes (the green dotted circle), the~lobe walls (the blue dotted lines), and~the relatively sharp cutoff of the dense torus at its outer~boundary (the outer red dotted lines). }
    \label{fig:density-example}
\end{figure}

Certain regions in our model have sharp boundaries. The~boundary is sharp between the lobe walls and the torus, between~the lobe and the lobe walls, and~between the ansae and the lobes. These boundaries are defined with straight lines and conic sections. Each ansa is defined as a uniform density elliptical region with sharp boundaries (see Figure~\ref{fig:density-example} in Section
~\ref{dust-distribution}). The~inner and outer boundaries of the lobe walls are conic section curves that become linear beyond a defined radius. The~parameters defining these boundaries are presented in Appendix \ref{model-parameters}. However, for~the purpose of simulating the radial cutoff of the torus discussed in Section~\ref{introduction}, the~boundary has the following continuous but relatively abrupt functional form:
\begin{equation}
    c_{out}(r)=0.5\times\left[1-\tanh{\left(\frac{r-R_{out}}{dr_{out}}\right)}\right].
    \label{eqn:cutoff-outer}
\end{equation}

This function decreases smoothly from 1 to 0, as~the radial distance to the central star, $r$, increases beyond $R_{out}$, with~$dr_{out}$ being the characteristic width of the cutoff zone. {The function multiplies the density obtained from the continuity equation to determine the final density.} $R_{out}$ and $dr_{out}$ are free parameters in our model-optimization process. Near~the star, an~{\it inner} cutoff function is imposed on the torus so that the density decreases smoothly to 0 as we approach the star: $c_{in}(r,\phi)=0.5\times(1-\tanh{\frac{R_{in}-r}{dr_{in}}})$.  We use prescribed parameters $R_{in}=2\times10^{14}$ cm (13.4 AU), and~$dr_{in}=2\times10^{13}$ cm, motivated by the results \linebreak of \citet{sahai_massive_2006} and \citet{maldoni_iras_2008} for the radial distances of dust formation~zones. 

\subsection{Radiative~Transfer} \label{radiative-transfer}

The only source of light in our model is the central star. Our radiative transfer calculation follows two successive photon paths, one from the star(s) to the scattering point, and~the other from the scattering point to earth. We calculate the optical depth along these paths accounting for dust density, albedo, and~extinction efficiency factors at different grain sizes, derived from the dust properties discussed in Section~\ref{dust-distribution}. Multiple scattering can make an additional contribution to the intensity arising from certain locations within the nebula, but~that is generally a small effect and would require a time-intensive, non-local approach, which we do not attempt here. Therefore, we make approximate higher-order scattering corrections to the points in the image that we judge are most affected by multiple scattering, as~we describe in Appendix \ref{app:radiative-transfer}. A~future extension of this study that would explicitly include multiple scattering could be performed by utilizing our derived model parameters with a Monte-Carlo procedure such as that described by \citet{yusef-zadeh_bipolar_1984}. 

After defining the density distribution and~executing the radiative transfer calculation, we convolve the final image with the point-spread function (PSF) in order to simulate the HST observations. The~PSF is extracted using two isolated point sources located to the Northeast and Southwest of M1-92 within the image. We take the average of these two profiles to reduce the effect of possible spatial PSF variability and intra-pixel position dependence \citep{mcmaster_hubble_2008}. Publicly available WFPC2 PSFs are not used since none of them share an identical set of observational parameters with our~data. 

\section{Characteristics} \label{characteristics}

In Table~\ref{tab:strongly-constrained-parameters}, we list model parameters inferred from strong observational constraints. These best-fitting values of the parameters are denoted by $^*$. The~best-fitting values are largely independent of altering other model parameters, except~for the ones listed as dependent in the table. We also assign uncertainty values to each parameter based on manual exploration of the tightness of their constraints. We designate these uncertainties as $\sigma_{Mtot}$, $\sigma_{\theta}$, and~$\sigma_{Rout}$, even though they are not statistically determined~quantities.

\begin{figure}[H]

\begin{adjustwidth}{-\extralength}{0cm}
\centering 
    \includegraphics[width = 6.7in]{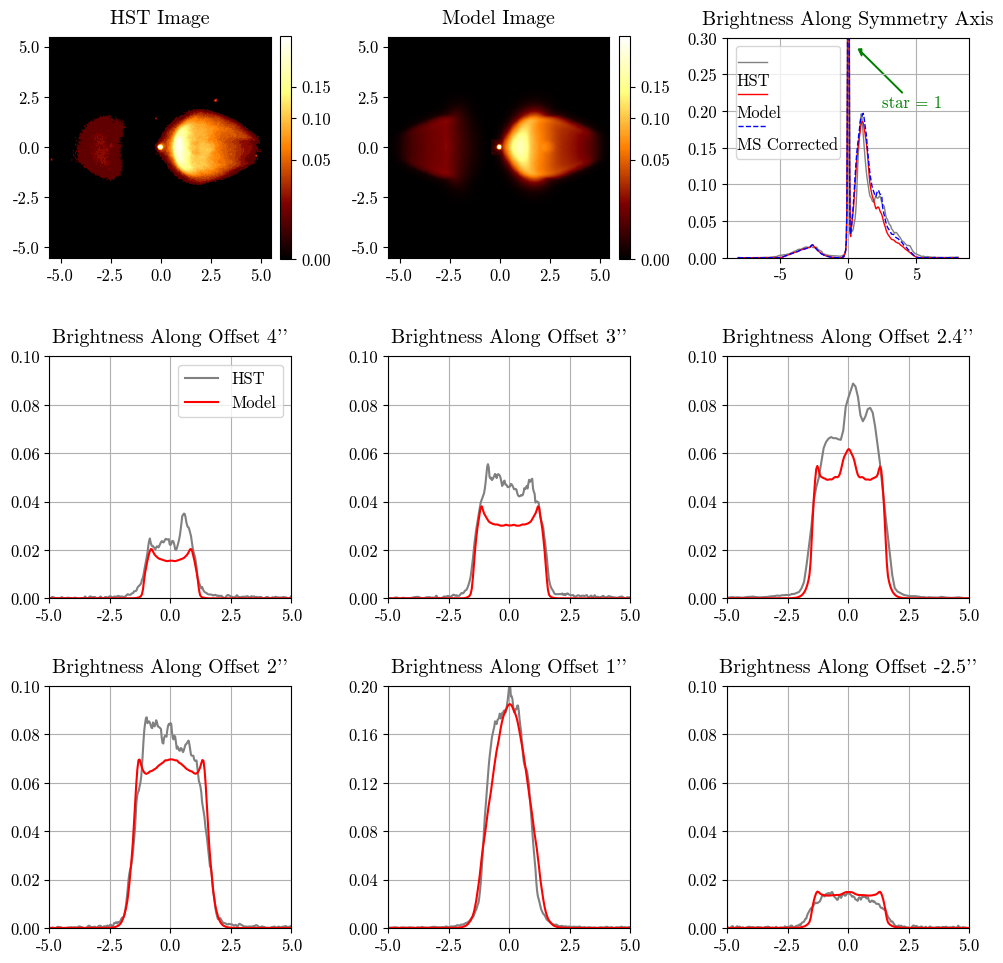}
\end{adjustwidth}
    \caption{Best-fitting 
 results for an inclination of $66^\circ$ using the density distribution of Figure~\ref{fig:density-example}. The~observed and simulated images in the upper panels are plotted with a square-root scale and~normalized so that the star's brightness is 1. All displayed profiles have the central star's brightness normalized to 1. The~brightness profile along the symmetry axis (upper right panel) shows a crude multiple-scattering-corrected curve (dashed), which is detailed in Appendix \ref{app:radiative-transfer}. Note: the scale on the lower middle panel differs from other profile~panels.}
    \label{fig:density-a}
\end{figure}

The sharpness of the outer cutoff of the torus is specified by the cutoff width parameter, $dr_{out}$, divided by the cutoff radius, $R_{out}$. Our parameter search showed that modeling results are insensitive to the assumed inner cutoff radius for the $66^\circ$ inclination model, for~$R_{in}$ values less than $2\times10^{16}$ cm. This is likely because the torus is thin at smaller radii, allowing almost all starlight to escape the immediate stellar environment to illuminate the reflection nebula, and~because the innermost portion of the torus does not absorb light along the observer's line of sight to the star. We subsequently use ``cutoff radius'' to refer only to the outer cutoff radius. The~best-fitting density distribution is illustrated in Figure~\ref{fig:density-example}, and~the corresponding simulated image and profiles are displayed in Figure~\ref{fig:density-a}. 
\vspace{-6pt}
\unskip

\section{Analysis} \label{analysis} 

\begin{figure}[H]

\begin{adjustwidth}{-\extralength}{0cm}
\centering 
 \includegraphics[width = 7in]{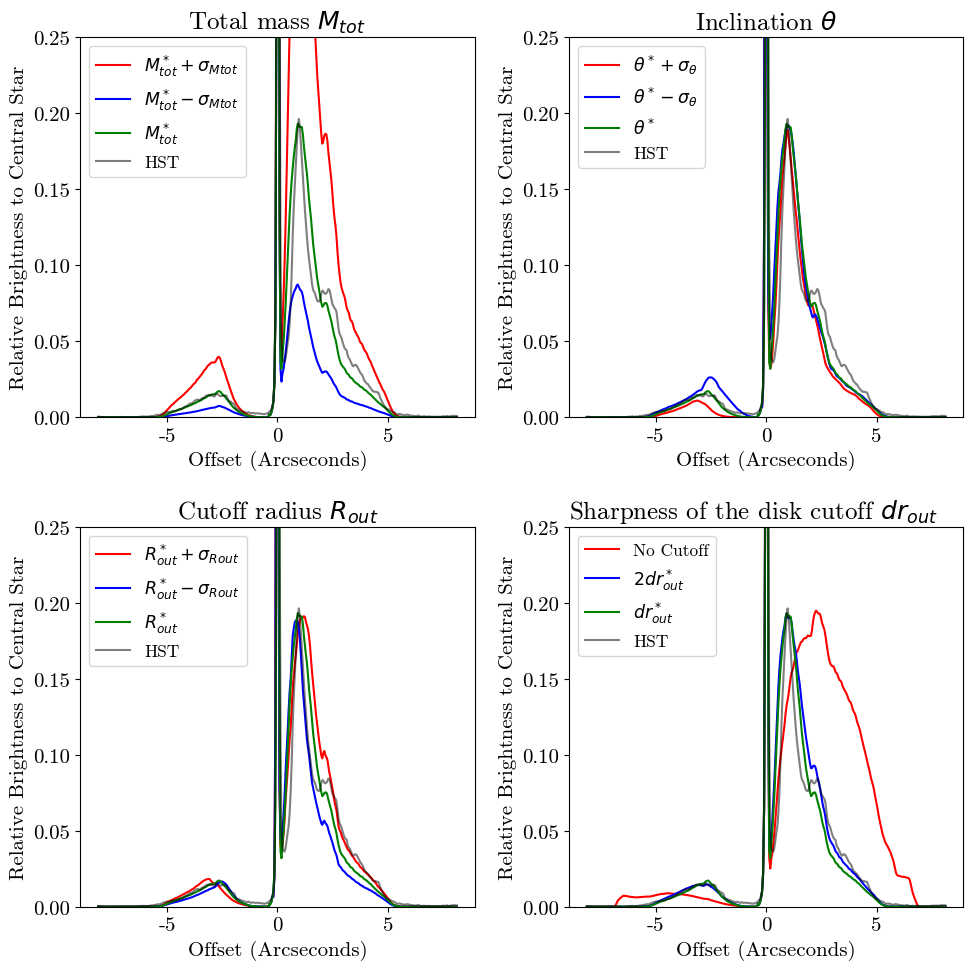}
\end{adjustwidth}

    \caption{Effects 
 of 
 changing each strongly constrained parameter on the intensity cut along the symmetry axis. \textbf{Top Left}: Changing the total mass by 1-$\sigma$ above or below the best estimated value. \textbf{Top Right}: Changing the inclination by 1-$\sigma$ above or below the best estimated value. \textbf{Bottom Left}: Changing the cutoff radius by 1-$\sigma$ above or below the best estimated value. \textbf{Bottom Right}: Increasing the cutoff characteristic~length.}
    \label{fig:parameter-uncertainty}
\end{figure}

As described in Section~\ref{characteristics}, the parameters listed in Table~\ref{tab:strongly-constrained-parameters} are highly sensitive to observational constraints. In~Figure~\ref{fig:parameter-uncertainty}, we show the effects of changing each parameter higher and lower by $\sigma$. The~plotted curves illustrate the effects of changing the parameters on the simulated light~profiles. 

Increasing the total mass while leaving the mass ratios of all the components constant directly boosts the lobe's brightness relative to the star. {As mass increases, dust density increases in both the lobes and the torus. Optical depth increases in the optically thick torus, decreasing the star's observed brightness. Scattering increases in the optically thin lobes, making the lobes brighter relative to the star. Decreasing the mass will conversely make the star brighter relative to the lobes.} This aligns with \citet{koning_post_2013}, who point out that increasing the total mass brightens the lobes and obscures the central region of the nebula. The~constrained total mass value is dependent on the inclination because~the constraint based on the star's brightness changes with the latitude of the line of~sight.

Through modeling experiments, we observe that the contrasting brightness of the two lobes on opposite sides of the system's midplane is ascribable to the inclination of the axis of symmetry. As~described by the scattering phase function (see Section~\ref{methods}), the~front/bright Northwest lobe points toward us and therefore has experienced a smaller scattering angle and consequently appears brighter than the rear southeast lobe, which points away from us. The~latter appears dimmer, in~part because its reflected light has undergone a large scattering angle, and~in addition, light from the base of the rear lobe is blocked by the opaque torus. As~shown in the top-right panel of Figure~\ref{fig:parameter-uncertainty}, decreasing the inclination increases the brightness contrast between the lobes (even beyond the torus cutoff), and~vice~versa. 

Although the density of the torus already decreases as the inverse square of the radial distance from the star, we show in Figure~\ref{fig:parameter-uncertainty} that an outer radial cutoff is still necessary to produce the observed absorption at the base of both lobes. As~seen in the observed images (Figure \ref{fig:hst-minkowski}), the~base of both lobes is obscured by the torus, the~base of the rear lobe to a much greater extent. In~our parameter experiments, we softened and even removed the outer cutoff, which caused the torus to extend further from the star. However, in~that case, with~the total mass simply constrained by the optical depth of the star, the~inferred density in the torus would decrease because the dust in the torus would occupy a larger volume. Furthermore, the~lobes would then appear brighter relative to the star. In~the case where there is no cutoff, optically thin regions of the torus at high radial distances can contribute to scattered light, increasing the extent of the~lobes. 

\section{Discussion} \label{discussion}

As shown in Figure~\ref{fig:density-example}, our model includes the lobe walls \citep{bujarrabal_direct_1994, bujarrabal_shock_1997, ramos-medina_spatio-kinematics_2014, alcolea_m_2022} and the \linebreak ansae \citep{soker_formation_1990, morris_bipolar_1990, trammell_hubble_1996}, and~constrains their shapes further through radiative transfer modeling. The~combination of a pair of diffuse, optically thin lobes, and~a relatively dense, constant-density lobe-wall structure that permits copious scattering is the only setup we found that produces the observed ``volume filled'' appearance of the lobes. This aligns with the results of \citet{koning_post_2013}. 

{\citet{koning_post_2013} employ isotropic scattering in their models. Using forward-dominant scattering, we find that the scattering phase function is the principal factor in producing the fainter rear lobe beyond the torus radial cutoff. Since the torus is optically thick and sharply cut off, it completely obscures the base of the rear lobe with little absorption beyond the cutoff.} \citet{koning_post_2013} show that in the general case, a~pair of bipolar cavities alone is sufficient to produce a torus-like appearance at the waist of a bipolar nebula. However, our model shows that lobe walls alone are not sufficient to reproduce the observed absorbed waist for M1-92. A~separate optically thick torus is required to extinguish direct starlight to match the observed star-to-lobe brightness~relationship. 

{The density law in the lobe interiors is not constrained in the final model, as~the interiors of the evacuated lobes are optically thin and scatter very little light. For~the lobe walls, a~uniform density law with respect to both radial distance and latitude best fits the observed light profile along the axis. The~thickness of the lobe walls is not constant and is constrained by the profiles perpendicular to the symmetry axis.}

{In the final constrained density distribution, the~torus has the shape of a flared, expanding disk that enfolds the lobes at larger radial distances. The~thin, disk-like portion has a radius of roughly 0.5'', and~its density and properties are loosely constrained.} The radius of the thin region in our model is consistent with the ``thin disk'' seen in the $J = 2 \rightarrow 1~^{13}\text{CO}$ observations performed by \citet{alcolea_minkowskis_2007}. Furthermore, both the torus and the lobe walls have densities higher than the critical density of the $^{13}\text{CO}~J = 2 \rightarrow 1$ line, while the lobe interiors have densities lower than the critical density. This is consistent with the images reported by \citet{alcolea_minkowskis_2007}, which show no $^{13}\text{CO}~J = 2 \rightarrow 1$ emission from the lobe~interiors. 

\subsection{Sharp Torus~Cutoff} \label{sharp-torus-cutoff}

\begin{figure}[H]

   \includegraphics[width = 5in]{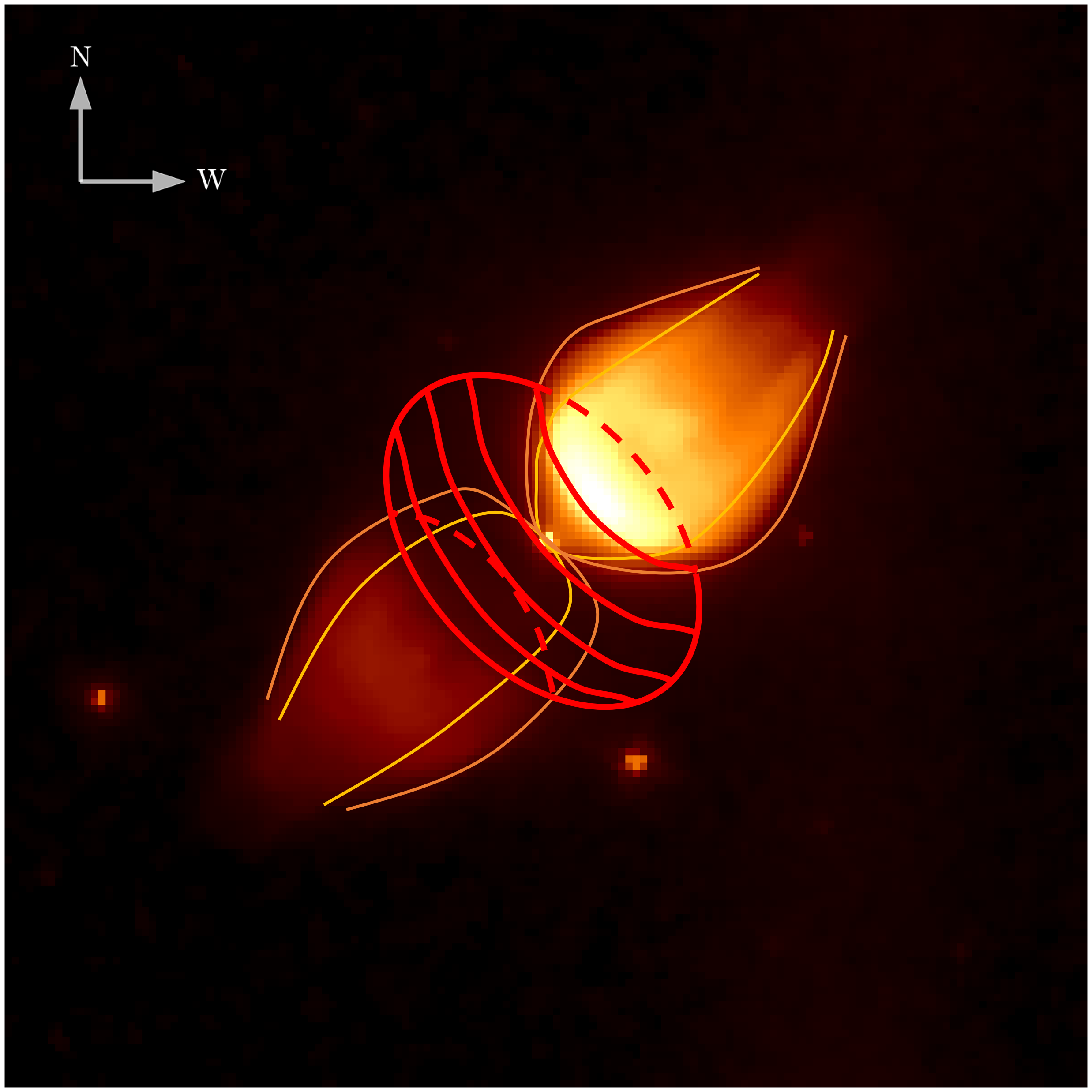}

    \caption{{How} the radial cutoff to the torus affects the appearance of the nebula. The~underlying {HST} image is displayed with a linear scale ranging from 0 to 20 counts per second. The~brown and orange lines represent roughly the illuminated lobe walls. The~red ellipse and its interior represent the torus, which covers a portion of both lobes. The~torus' high optical depth obscures almost all scattered light, and only some portion of the direct starlight is able to penetrate it. \textit{In the observed images, the~transition between the lobes and the absorbed region is sharp, thereby signifying the sharp outer cutoff to the torus. 
}}
    \label{fig:structure-appearance}
\end{figure}

The existence of the sharp radial cutoff to the torus manifests predominantly in the light distribution of the lobes. This is illustrated in Figure~\ref{fig:structure-appearance}. As~shown on the bottom right plot of Figure~\ref{fig:parameter-uncertainty}, the~radial cutoff is responsible for producing the sharp transition from the completely obscured region near the central star to the peak of the rear lobe. The~sharp cutoff leads to a rapid brightness increase at a radius of 0.5'' to 1'' in the near lobe. As~shown in the top-left panel of Figure~\ref{fig:density-a}, the~transition between the optically thin and optically thick regime is rapid. In~the optically thick region of the torus (with an angular offset along the symmetry axis between 0.5$''$ to $-$2$''$), only the light from the star is able to effectively penetrate the absorbing~dust. 

Sharp boundaries are not uniquely identified in M1-92; they are also observed in other PPNe. Molecular-line observations of the ``water-fountain'' PPN, IRAS 16342$-$3814, show the presence of a similarly sharply-truncated expanding molecular envelope \citep{sahai17}. \citet{sahai_discovery_1999} and \citet{kwok_discovery_2000} have identified other preplanetary nebulae with sharp outer boundaries in their equatorial regions, as seen in their optical images. Examples include CRL 2688 (Egg Nebula) \citep{sahai_imaging_1998, balick_illumination_2012}, Hen 3-
1475 (IRAS 17423$-$1755) \citep{sahai_preplanetary_2007}, IRAS 19292$+$1806 \citep{sahai_preplanetary_2007}, IRAS 22036$+$5306 \citep{sahai_icy_2003, sahai_preplanetary_2007}, and~OH 231.8$+$4.2 (Calabash Nebula/Rotten Egg Nebula) \citep{bujarrabal_hst_2002}. 




\subsection{Inclination} \label{inclination}

With an assumed scattering asymmetry factor $g=0.6$, we infer that the scattered light image of M1-92 leads to a preferred inclination of $\theta^*(g=0.6) = 66^\circ \pm 5^\circ$. This value is somewhat different than previously suggested values: $60^\circ$ \citep{herbig_spectrum_1975}, $55^\circ$ \citep{davis_oh_1979}, \mbox{$57^\circ \pm 5^\circ$ \citep{solf_long-slit_1994}}, $51.5^\circ$ \citep{alcolea_minkowskis_2007}, $52^\circ$ \citep{alcolea_new_2008}, and~finally $50^\circ \pm 5^\circ$ \citep{alcolea_m_2022}. These estimates mainly originate from velocity measurements. Since velocity measurements are radial, these estimates utilize assumptions that connect the radial velocity to the three-dimensional expansion velocity in the nebula. {For example, assuming that the lobes and the torus have the same kinetic age,\linebreak  \citet{alcolea_minkowskis_2007, alcolea_m_2022} estimate M1-92's inclination to be $51.5^\circ$.} For our inferred $\theta^*=66^\circ$, the~kinetic age inferred for the torus using radial velocity measurements will increase, implying that the torus formed prior to the bipolar~lobes. 

\begin{figure}[H]

\begin{adjustwidth}{-\extralength}{0cm}
\centering 
    \includegraphics[width = 7in]{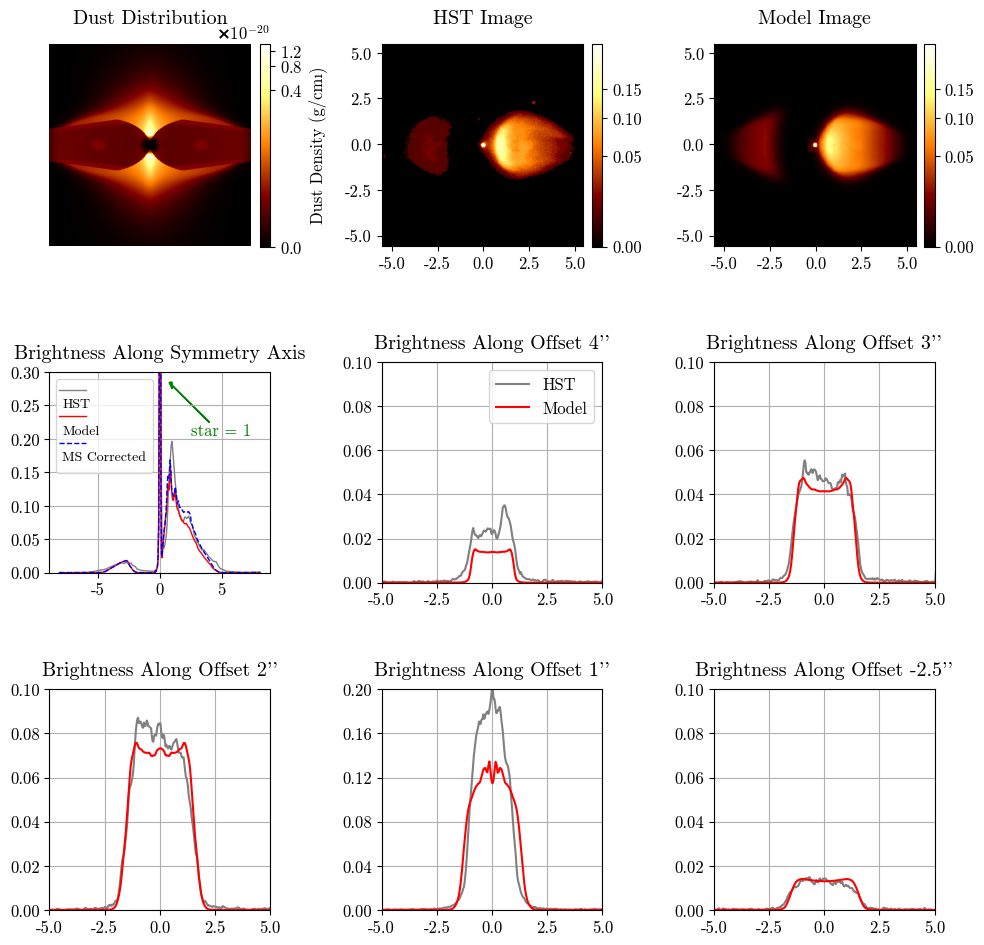}
\end{adjustwidth}

    \caption{Best 
 fitting 
 results for an inclination of $55^\circ$. The~\textbf{top-left} panel shows the density distribution in a meridinal cross-section, with~values scaled to the  power of $0.2$. The~\textbf{top-middle} and \textbf{top-right} panels show images in a square-root scale. Note the strong latitude-dependence and the elongated~torus. }
    \label{fig:density-b}

\end{figure}

We have considered an alternative in which $\theta=55^\circ$, as shown in Figure~\ref{fig:density-b}. We have gone through the same optimization process as described in Section~\ref{methods} and altered all parameters for this $55^\circ$ model, in~which case the scattering asymmetry factor is constrained to $g=0.35\pm0.07$. Young grains forming out of the material being ejected from the central object of M1-92 could possibly be smaller, on~average, than~standard interstellar dust. If~so, then from the dust grain studies presented by \citet{weingartner_dust_2001} and\linebreak  \citet{draine_scattering_2003}, such grains could exhibit a smaller scattering asymmetry factor. On~the other hand, \citet{murakawa_evidence_2010} suggest that the grains in M1-92, excluding the torus, have a size distribution typical of those found in the interstellar medium. (Grains in the torus can grow to even larger sizes up to $1000\;\upmu$m, due to coagulation that follows grain--grain\linebreak  collisions \citep{chokshi_dust_1993, murakawa_evidence_2010}.) Thus, a~scattering asymmetry factor of 0.6 derived from interstellar dust and reflection nebulae seems more plausible \citep{weingartner_dust_2001, draine_interstellar_2003, witt_scattering_1982, witt_scattering_1990}. Furthermore, our model with a $66^\circ$ inclination additionally performs slightly better, with~a weighted RMS deviation of 0.017 as opposed to 0.019 in the $55^\circ$ model. 

The higher weighted RMS deviation of the $55^\circ$ model is most likely attributable to an inherent preference for a larger inclination when working with the optical evidence. As~shown in Figure~\ref{fig:structure-appearance}, the~inferred torus must be optically thick at systemic latitudes higher than the observing latitude to obscure the base of the bright near lobe. In~the model with an inclination of $66^\circ$, the~remaining latitude space occupied by the lobes matches the observed lobe angular width in the image. With~a lower assumed inclination, the~optically thick torus would need to extend to higher latitudes; consequently, the lobes would have to be thinner and~occupy a narrower range of high latitudes. Furthermore, an~extension of the torus to higher latitudes implies that the change in optical depth with latitude is more drastic, suggesting a stronger latitude dependence in the mass loss, which does not align as well with the observed scattered light in the lobes far from the star (where there is minimal absorption), as~can be seen by comparing Figure~\ref{fig:density-a} to Figure~\ref{fig:density-b}.

As discussed in Section~\ref{analysis}, our model constrains the inclination $\theta$ of the symmetry axis using the observed brightness ratio between the brightest pixel on the front lobe and the brightest pixel on the rear lobe (front-to-rear brightness ratio) and the asymmetry factor $g$. This three-way correlation between the front-to-rear brightness ratio, $g$, and~$\theta$ can be observed in similar objects. In~the HST WFPC2 F606W image of the preplanetary nebula CRL 2688 \citep{sahai_bipolar_2002}, the~front-to-rear brightness ratio is about 5, lower than in M1-92. By~the relationship discussed in Section~\ref{analysis}, CRL 2688 should have a higher inclination. This aligns with \citet{sahai_structure_1997}'s estimated inclination of $70^\circ$ to $80^\circ$. Another example that shows this relationship is the preplanetary nebula Hen 3-1475. \citet{borkowski_kinematics_2001} estimate Hen 3-1475's inclination to be $40^\circ$. Compared to M1-92, this lower inclination value should increase the front-to-rear brightness ratio. The~prediction aligns with the HST WFC3 F555W observation, which shows a front-to-rear brightness ratio of over 20 \citep{balick_watching_2011}. In~conclusion, the~correlation between $\theta$ and the front-to-rear brightness ratio is not unique to M1-92 and can be identified in similar~objects.

\section{Summary and~Conclusions} \label{conclusion}

In this study, we present a dust-scattering model of M1-92 that reproduces the observed Hubble Space Telescope (HST) F547M image. We have derived parameters of the shape, size, and~mass of commonly suggested components of M1-92. {Our total mass estimate of $1.10 \pm 0.1\;M_\odot \left(\frac{d}{2.5\text{kpc}}\right)^2  \left(\frac{\text{GDR}}{200}\right)$ is consistent with the previously estimated $0.9\;M_{\odot}$ \citep{alcolea_minkowskis_2007, alcolea_new_2008}. Although~the density law within the lobe interiors is unconstrained, the~best fitting value for the lobe mass loss rate parameter $6.25\times 10^{-7}\;M_{\odot}/yr$ fits within the range typically assumed for the mass loss rate of fast winds, i.e.,~$10^{-8}$ to $10^{-6}\;M_{\odot}/yr$ \citep{balick_shapes_2002, lee_shaping_2003}.}

{From the optical image, we show that the torus exhibits a radial density cutoff at large radii. We propose that the presence of sharp outer boundaries of tori/disks in preplanetary nebulae is caused by the rapid onset of mass ejection resulting from a strong binary interaction, such as the relatively sudden onset of CE evolution. M1-92 is not unique in this regard: \citet{sahai17} infer that the ``water-fountain'' PPN, IRAS 16342$-$3814, has resulted from CE ejection. The~other objects observed with sharp outer boundaries listed in Section~\ref{sharp-torus-cutoff} might have undergone similar binary-companion-driven processes during their formation. M1-92's central object has long been known as an evolved binary system from UV spectroscopy \citep{feibelman_ultraviolet_1990}.} 


{One possible mechanism for material ejection during common-envelope evolution has been suggested by \citet{glanz_efficient_2018}. In~this scenario, the~inspiral of the companion in a common envelope induces the expansion of the stellar envelope, which then cools, allowing dust condensation. Dust is then accelerated by the radiation pressure of the star, and~eventually results in the entire envelope being ejected through collisional coupling to the gas. The~expanded, massive torus in M1-92 implies that most of the common envelope has been ejected, and~its central star and companion have moved beyond the common envelope configuration. High-velocity mass loss in the lobes could be driven by mechanisms such as a differentially rotating accretion disk around the core of the AGB star, producing collimated polar streams \citep{morris_bipolar_1990, sahai_collimated_2003}. In~this case, the~torus formed before the initiation of the collimated outflows as the stellar interior becomes exposed \citep{morris_mechanisms_1987, morris_bipolar_1990, soker_formation_1990, soker_jet_1992}, and~the collimated outflows sculpt and compress the torus from the inside out, forming the lobe walls and the evacuated lobe interiors. Such a mechanism within which high-speed collimated outflows carve out the lobes within a pre-existing envelope was proposed by \citet{sahai_multipolar_1998}.} 


{Several studies have investigated the interactions between collimated fast winds (CFWs) and pre-existing envelopes. These findings are informative and relevant, though~the sources of the CFWs in these studies are not likely identical to the source in M1-92.\linebreak  \citet{soker_formation_2002} studied the interaction between CFWs blown by compact accreting companions and a pre-existing AGB wind. He concluded that the CFW produces an axisymmetric bubble within the slow wind, which is narrow near and far from the central star(s) and~wider in between. \citet{lee_shaping_2003} performed simulations of tapered CFWs reshaping a pre-existing AGB wind, which produced bipolar dense-walled lobes with tenuous interiors, as~we find for M1-92. The~observed increase in lobe wall thickness with radial distance along the symmetry axis in our models is consistent with the predictions of the hydrodynamical models presented in \citet{lee_shaping_2003}. \citet{balick_models_2019} tested larger taper opening angles in their models, producing lobes with shapes and structures near their bases (i.e., relatively close to the central star) that are qualitatively consistent with the density distribution that we infer in our study.}


{An alternative possibility for producing the collimated fast outflow is presented by \citet{icke_cylindrical_2022}, in~which an equatorial gaseous toroidal ring is photoevaporated by the star and~produces a wind that interacts with a constant density ambient medium. Hydrodynamical simulations show that in such a scenario, one can obtain a bipolar nebula with pronounced cylindrical shaped lobes, such as those observed in the PPN Hen 3-401. Such a photoevaporative flow model may be applicable for M1-92 as well.} 



{Another possible mechanism for material ejection was suggested by \citet{alcolea_minkowskis_2007}, who proposed that a magneto-rotational explosion simultaneously initiated the bipolar jets and the equatorial flows. The~sudden initiation of mass loss would explain the sharp cutoff to the torus. However, such a formation mechanism would require the kinetic age of the torus to match the kinetic age of the lobes. As~discussed in Section~\ref{inclination}, \citet{alcolea_minkowskis_2007} point out that if the nebula's inclination is $51.5^\circ$, radial velocity measurements would indicate a common kinetic age of $1200$ yr for the torus and the lobes. Our inclination estimate of $66^\circ$ implies the torus formed prior to the bipolar lobes, given the radial velocity measurements, which is not consistent with the magneto-rotational explosion model.}

The tension between the inclinations derived from scattered light and velocity measurements, {which lead to different implicated formation schemes, }can be addressed in the future by combining the two approaches and~by elaborating dust-scattering models using observations at multiple wavelengths and by employing a multiple-scattering radiative transfer model. The additional refinement of the dust distribution in M1-92 can also be pursued using polarization observations and observations in the thermal infrared, combined with models that account for both the scattered and thermal dust emissions 
  \citep{murakawa_evidence_2010}.

\vspace{+6pt}
\authorcontributions{Conceptualization, M.R.M.; formal analysis, Y.Q.L.; investigation, Y.Q.L. and M.R.M.; methodology, Y.Q.L. and M.R.M.; software, Y.Q.L.; supervision, M.R.M.; validation, Y.Q.L., M.R.M., and R.S.; writing---original draft preparation, Y.Q.L., M.R.M., and R.S.; writing---review and editing, Y.Q.L., M.R.M., and R.S. All authors have read and agreed to the published version of the manuscript.}

\funding{This research received no external funding.}

\dataavailability{The data described here may be obtained from the MAST archive at
\url{https://dx.doi.org/10.17909/awd4-de67}.}

\acknowledgments{This research is based on observations made with the NASA/ESA Hubble Space Telescope obtained from the Space Telescope Science Institute, which is operated by the Association of Universities for Research in Astronomy, Inc., under NASA contract NAS 5–26555. These observations are associated with programs 6533 and 6761. 
We are grateful to the anonymous referees and editors for their valuable comments, helpful suggestions, and~kind words. Notably, their dedication and time commitment greatly contributed to the development of this~manuscript.
R.S.’s contribution to the research described here was carried out at the Jet Propulsion Laboratory, California Institute of Technology, under~a contract with NASA, and~funded in part by NASA via ADAP awards and~multiple HST GO awards from the Space Telescope Science~Institute.}

\conflictsofinterest{The authors declare no conflict of interest.}

\appendixtitles{yes} 
\appendixstart
\appendix

\section{Parameters} \label{model-parameters}
\unskip

\subsection{Fixed~Parameters}

The parameters in the following table are adjustable in our model; however, for~the purpose of this study, they are kept at a constant~value. 

\begin{table}[H]

\caption{Fixed~parameters}
\label{tab:parameter-fixed}
\setlength{\tabcolsep}{20.75mm}

\begin{tabular}{@{}lll@{}} 
\toprule
\textbf{Parameter Definition} & \textbf{Value} \\
\midrule
Observing wavelength & $0.547\;\upmu$m \\
Observer distance to nebula & $2.5$ kpc \\
Gas-to-dust ratio & $200$ \\
Dust grain minimum size & $5\times10^{-3}\;\upmu$m \\
Non-torus dust grain maximum size & $0.5\;\upmu$m \\
Torus dust grain maximum size & $1000\;\upmu$m \\
Dust grain density & $3\text{ g/cm}^3$ \\
Lobe interior outflow velocity & $500$ km s$^{-1}$ \\
Torus outflow velocity & $10$ km s$^{-1}$ \\
\bottomrule
\end{tabular}
\end{table}
\unskip

\subsection{Constrained~Parameters}

All the parameters listed in Table~\ref{tab:parameter-best} are constrained by the model. Unlike the  values listed in Table~\ref{tab:strongly-constrained-parameters} or Figure~\ref{fig:density-example}, they do not represent~measurements. 

The torus/lobe mass loss rate parameters are used in the continuity equation to calculate the dust density within the torus and the lobes. The~``Mass loss latitude dependence'' $a_m$ parameter specifies an additional latitude dependence to the mass loss rate, where $1$ indicates no additional latitude dependence  and~$8$ indicates an additional latitude dependence 8 times stronger in the equatorial plane than in the axis of symmetry\endnote{This factor is calculated as an axisymmetric spheroid with a semi-major axis of $2$ and a  semi-minor axis of $1/4$ and~superimposed onto the other density calculations.}. 


``Location of the ansa's base'' is the distance between the star(s) and the closest point on the ansae, and~``Location of the ansa's focus'' is the distance between the ansae's focus and the star(s). ``Lobe/lobe wall outer cutoff distance'' and ``Sharpness of the lobe/lobe wall outer cutoff'' are represented by $R_{out}$ and $dr_{out}$ for both the lobes and the lobe walls. As~described in the main text, the~shape of the boundaries between the lobe walls and the lobes and between the lobe walls and the torus are conic section curves that transition to linear beyond a distance from the base of the conic section defined by parameter ``Transition distance'' $D_{trans}$. ``Base location'' $x_{base}$ specifies where the boundary intersects the axis of symmetry. ``Conic section eccentricity'' $e$ defines the eccentricity of the conic section region. ``Focus location'' $f$ specifies the distance from the star along the axis of symmetry of the focus of the conic section curves. ``Slope of the linear boundary'' $a$ represents the slope of the linear section of the lobe wall boundaries. Given the above parameters, the~position of the directrix $l$ of the conic section region relative to the star can be calculated as
\begin{equation}
l = -\frac{f + x_{base}}{e} - x_{base}
\label{eq:directrix}
\end{equation}

We then obtain several supplementary parameter values, which will allow us to define the boundary in the meridional plane, with~$x$ being the position along the axis of symmetry and~$y$ being the vertical separation from the boundary to the axis of symmetry along the specified $x$. The~supplementary parameters are the meridional plane position of the transition point between the conic section and the linear region $(x_{cut}, y_{cut})$, as~well as the $y$ intercept of the line in the meridional plane that describes the linear region of the boundary $y_{intercept}$. We numerically find these values that satisfy our defined $D_{trans}$, and~$y$ can thus be expressed as 


\begin{equation}
y(x) =
\begin{cases} 
\sqrt{\begin{array}{l}
(e \cdot (x - l))^2 - x^2
- f^2 + 2 \cdot x \cdot f
\end{array}} & \text{if } x \leq x_{\text{cut}} \\
y_{\text{intercept}} - x \times a & \text{otherwise}
\end{cases}
\label{eq:y}
\end{equation}

Since the lobes are narrow, with~a measurable value of $y$, we can estimate $\hat{A}(r,\phi)$ as the area of the circle with radius $y$:
\begin{equation}
\hat{A}(r,\phi)=4\pi y^2
\label{eq:ahat-lobes}
\end{equation}

For the torus, we define $\hat{A}(r,\phi)$ as the area of a sphere with radius $r$ beyond a radial distance of $1.8 \times 10^{16} \text{ cm}$. Since the torus is compressed near the star, within~$1.8 \times 10^{16} \text{ cm}$, the density scales as $r^{-1}$.
\begin{equation}
\hat{A}(r,\phi) =
\begin{cases} 
4\pi r^2 & \text{if } r \geq 1.8 \times 10^{16} \text{ cm} \\
4\pi r \cdot (1.8 \times 10^{16} \text{ cm}) & \text{if } r < 1.8 \times 10^{16} \text{ cm}
\end{cases}
\label{eq:ahat-torus}
\end{equation}

With these expressions, the~``Torus mass loss rate'' parameter is the extrapolated mass loss rate if it were isotropic. We can obtain $\dot{M}(r,\phi)$ by

$$\dot{M}(r,\phi)=\text{Torus mass loss rate}\cdot(\frac{1}{2}-\frac{r}{2\sqrt{r^2+y(r)^2}})\cdot\sqrt{a_m^2\cos^2\phi+\frac{1}{a_m^4}sin^2\phi}$$

\begin{table}[H]
\caption{Constrained parameters for both~models}
\label{tab:parameter-best}
\setlength{\tabcolsep}{0.95mm}

\begin{adjustwidth}{-\extralength}{0cm}
\begin{tabular}{@{}lll@{}} 
\toprule
\textbf{Parameter Definition} & $\mathbold{66^\circ}$ \textbf{Inclination Model} & $\mathbold{55^\circ}$ \textbf{Inclination Model} \\
\midrule
Torus mass loss rate & $6.27\times10^{-4}~M_{\odot}/$year & $7.98\times10^{-4}~M_{\odot}/$year \\
Lobe mass loss rate & $6.25\times10^{-7}~M_{\odot}/$year & $7.95\times10^{-7}~M_{\odot}/$year \\
Mass loss latitude dependence & $1$ 
 & $8$\\
Asymmetry factor & $0.6$ & $0.35$ \\
Inclination & $66^\circ$ & $55^\circ$ \\
Lobe wall dust density (constant) & $9.88\times10^{-22}$ g/cm$^3$ & $3.12\times10^{-22}$ g/cm$^3$ \\
Ansae dust density (constant) & $3.95\times10^{-22}$ g/cm$^3$ & $2.52\times10^{-22}$ g/cm$^3$ \\
Conic section eccentricity---lobe wall-torus boundary& $0.85$ & $0.83$ \\
Conic section eccentricity---lobe-lobe wall boundary & $0.8$ & $0.83$ \\
Base location---lobe wall-torus boundary & $2.4\times10^{15}$ cm enclosing star & $0$ cm intersecting star \\
Base location---lobe-lobe wall boundary & $1.2\times10^{15}$ cm enclosing star & $6\times10^{15}$ cm enclosing star \\
Focus location---lobe wall-torus boundary & $1.38\times10^{16}$ cm & $1.68\times10^{16}$ cm  \\
Focus location---lobe-lobe wall boundary & $1.5\times10^{16}$ cm & $1.08\times10^{16}$ cm \\
Transition distance---lobe wall-torus boundary & $1.62\times10^{17}$ cm & $8.4\times10^{16}$ cm \\
Transition distance---lobe--lobe wall boundary & $1.26\times10^{17}$ cm & $8.4\times10^{16}$ cm \\
Slope of the linear boundary between lobe wall andtorus & $0.34$ & $0.145$ \\
Slope of the linear boundary between lobe and lobe wall & $0.32$ & $0.116$ \\
Location of the ansa's base & $8.4\times10^{16}$ cm from the star & $9\times10^{16}$ cm from the star \\
Location of the ansa's focus & $9\times10^{16}$ cm from the star & $9.6\times10^{16}$ cm from the star \\
Lobe/lobe wall outer cutoff distance & $1.8\times10^{17}$ cm & $1.9\times10^{17}$ cm \\
Sharpness of the lobe/lobe wall outer cutoff & $0.056$ & $0.053$ \\
Torus inner cutoff distance & $2\times10^{14}$ cm & $2\times10^{16}$ cm \\
Torus outer cutoff distance & $1\times10^{17}$ cm & $7\times10^{16}$ cm to $2\times10^{17}$ cm, elliptical \\
Sharpness of the torus outer cutoff & $0.02 \times (1 + 9 \times \cos^2{\phi})$ & $0.0286 \times (1 + 10 \times \cos^2{\phi})$ \\
\bottomrule
\end{tabular}
\end{adjustwidth}
\end{table}
\unskip

\section{Model~Details} \label{model-details}
\unskip

\subsection{Dust~Properties} \label{dust-properties}

The total scattering or extinction cross-section per unit volume is equal to the local mass density times the cross-section per unit mass, assuming the grains are statistically homogeneous. The~cross-section per unit mass is calculated by integrating the scattering cross-section of dust grains of each size with their relative abundance. Recall that we employ the Mathis--Rumpl--Nordsieck dust grain size distribution with $n(a)\propto a^{-3.5}$ \citep{mathis_size_1977, draine_physics_2011}. We can define $n(a)=k a^{-3.5}$ as the differential number density for grains of size between $a$ and $a+da$, where $k$ is a constant. For~a local dust mass density P, we can calculate $n(a)$ as 

\begin{equation}
    n(a)=\frac{3P}{8\pi \rho (a_{max}^{0.5} - a_{min}^{0.5})}a^{-3.5}
    \label{eqn:abundances-formula}
\end{equation}
\textls[-15]{where $a_{min}$ and $a_{max}$ are, respectively, the minimum and maximum grain size cutoff, and~$\rho$ is the assumed individual dust grain mass density, which we set equal to  $3$ g~cm$^{-3}$ \citep{draine_physics_2011} p. 245.} 


Therefore, the~scattering and extinction cross-section densities are calculated as
\begin{equation}
    C_{sca}=\int_{a_{min}}^{a_{max}} n(a)\pi a^2 Q_{sca}(\frac{2\pi a}{\lambda})da
    \label{eqn:scattering-cross-section-formula}
\end{equation}
\begin{equation}
    C_{ext}=\int_{a_{min}}^{a_{max}} n(a)\pi a^2 Q_{ext}(\frac{2\pi a}{\lambda})da
    \label{eqn:extinction-cross-section-formula}
\end{equation}

We use the scattering and extinction efficiency factor functions provided by \linebreak \citet{draine_physics_2011}, with~an assumed refractive index of $2+i$. In~the code, these integrals are performed partially numerically to~account for the non-analytic shape of the scattering and extinction efficiency factor functions:
\begin{equation}
    C_{sca}=\frac{3}{8\rho(a_{max}^{0.5} - a_{min}^{0.5})}\times P \times\int_{a_{min}}^{a_{max}} a^{-1.5}Q_{sca}(\frac{2\pi a}{\lambda})da
    \label{eqn:scattering-cross-section-results}
\end{equation}
\begin{equation}
    C_{ext}=\frac{3}{8\rho(a_{max}^{0.5} - a_{min}^{0.5})}\times P \times\int_{a_{min}}^{a_{max}} a^{-1.5}Q_{ext}(\frac{2\pi a}{\lambda})da
    \label{eqn:extinction-cross-section-results}
\end{equation}

\begin{figure}[H]
    \includegraphics[width = 8.8cm, angle = 0]{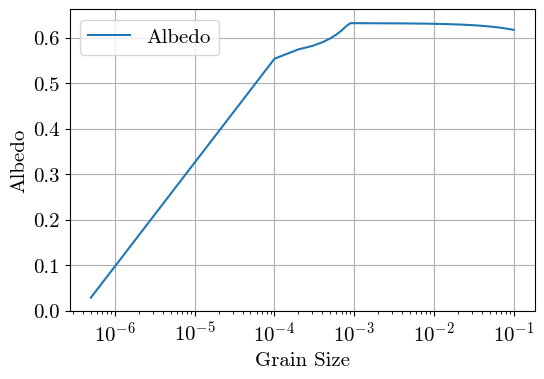}
    \caption{Albedo of the dust grains vs. grain~size. }
    \label{fig:dust-albedo}
\end{figure}

On the right side of the equals sign, the~expressions on the left are denoted as the torus/non-torus cross-section constants in our model, while the integral on the right is evaluated numerically and denoted as the scattering/absorption efficiency integral. These are separately evaluated for the torus and non-torus regions, where the maximum grain size is different, as presented in Section~\ref{dust-distribution}. $P$ is the local mass density, which we obtain from the continuity equation. Figure~\ref{fig:dust-albedo} displays the albedo for the dust grains as a function of~size. 

\unskip

\subsection{Radiative Transfer and Multiple~Scattering} \label{app:radiative-transfer}

For the radiative transfer calculation, we take advantage of the axial symmetry of the problem. We first create a two-dimensional function specifying the intensity of starlight reaching all points defined by radius, $r$, and~latitude $\phi$. This involves accounting for the inverse square law of the isotropic light source and~using the assumed dust density distribution to calculate the extinction along the line of sight from the star to each point specified by $(r, \phi)$. 

For a chosen observing latitude, and~for each pixel of the observer's projected image of the nebula, we integrate the intensity along the line of sight from the most distant extent of the computational domain to the nearest one (relative to the observer), taking into account the intensity of the starlight reaching each point $(r, \phi)$ along the line of sight, the~extinction from the point to the observer, the~scattering cross-section per unit volume at the scattering point, and~the scattering phase function. The~full 2D image of the nebula is then compiled for all pixels. The~star is treated as a point source, with~the optical depth toward the observer taken into consideration. At~last, we convolve the simulated image with the observational point spread function (PSF). 

The contribution of multiple scattering can be important. We assume that most of the multiple scattering relay points are in the lobe walls, since the lobe walls are also where singly scattered light is the most prominent. Because~forward scattering is much more prominent than backward scattering, a~final scattering point on the lobe walls will ``see'' most scattered light from relay points closer to the star, mostly lobe wall locations between the final scattering point and the star. Therefore, for~a final scattering point, the~multiple scattering intensity is estimated as follows: the total solid angle occupied by the lobe wall section between the scattering point and the star divided by $4\pi$ steradians, multiplied by the estimated intensity emerging from the lobe walls within that solid angle (the average singly-scattered intensity at the specified point). This is then multiplied to the estimated average absorption opacity for scattering off the lobe walls to obtain the estimated multiple scattering intensity. This is a crude estimation because no integration over the phase function is performed and~thus should be further verified using a multiple scattering~model. 

Based on this double-scattering correction, the~following predictions can be made: closer than 2$''$ from the star, we estimate that multiple scattering is not prominent, since the lobe wall section between these locations and the star occupy a very small range of solid angles. Approaching larger radii, the~lobe walls begin occupying larger angular areas. At~the same time, the~scattering angles at the relay points also decrease. Therefore, multiple scattering can become prominent at these~distances. 

\subsection{Finding Optimized~Parameters} \label{app:parameter-optimization}

As described in Section~\ref{methods}, the~optimal values for the four parameters listed in Table~\ref{tab:strongly-constrained-parameters} were found using optimization algorithms. In~this procedure, we first estimate the mass-loss rate parameter. As~previously discussed, the~ratio of the lobe brightness to the star brightness increases with the mass-loss rate, and~vice~versa. (Specifically, this ratio is determined by dividing the value of the brightest pixel on the front lobe by the brightest pixel of the star.)  This positive correlation allowed us to estimate the mass loss rate using a binary search: we started with a mass loss rate value of $3.2\times~10^{-4}M_{\odot}/$yr and an increment interval of $1.6\times~10^{-4}M_{\odot}/$yr. We increased the total mass by the increment interval if the brightness ratio between the front lobe peak and the star was smaller than the observed brightness ratio  and~vice~versa. Finding the mass loss rate value  that  yielded a brightness ratio discrepancy of less than $10\%$ between the model and the observation was sufficient for the initial estimation, while the final step required the discrepancy to be less than $1\%$.

To find the optimal inclination value, we used the brightness ratio between the lobes. Specifically, we calculated the ratio between the brightest pixel on the front lobe and the brightest pixel on the rear lobe. The~asymmetry factor was fixed at 0.6 in this process. We conducted a grid search over all integer angle inclination values to find the best fit. To~reduce computing time, a~range of inclinations that were too small or too large was manually excluded before the grid~search.

To find the cutoff radius and cutoff sharpness, we searched over a 2D grid of possible combinations. The~grid contained $R_{out}$ values between $5\times10^{16}$ cm and \mbox{$1.5\times10^{17}$ cm} (with increments of $1\times10^{16}$ cm), and~$dr_{out}$ values between $0$ cm and $1\times10^{17}$ cm (with increments of $1\times10^{16}$ cm). For~each combination, we calculated the weighted root-mean-square (RMS) deviation. This weighted RMS deviation is calculated by weighting the square deviation of each pixel by the pixel's observed value to the power of 0.5 (to balance the constraining strength of the front and rear lobes), then taking the square root of the summed and weighted average. After~finding the optimal values for both parameters, latitude dependence was introduced to both parameters if appropriate. The~introduced latitude dependence kept the optimized parameter values at a zero latitude. Finally, we executed the binary search again to find a precise mass loss rate estimate and integrated over the nebular medium to obtain the total~mass.

\begin{adjustwidth}{-\extralength}{0cm}

\printendnotes[custom]
\reftitle{References
}


\PublishersNote{}
\end{adjustwidth}

\end{document}